\documentstyle[12pt]{article}
\begin{document}
\title{The Euler Equations and Non-Local Conservative Riccati
Equations}
\author{Peter Constantin
\\Department  of Mathematics\\The University of Chicago }
\maketitle
\newtheorem{thm}{Theorem}
\newtheorem{prop}{Proposition}
The purpose of this brief note is to present an infinite
dimensional family of exact solutions of the incompressible three-dimensional
Euler equations
\begin{equation}
\partial_t{\mathbf u} + {\mathbf u}\cdot\nabla{\mathbf u} + \nabla p = 0,\quad
\nabla\cdot{\mathbf u} = 0.
\label{euler}
\end{equation}
The solutions we present have infinite kinetic energy and
blow up in finite time. Blow up of other similar infinite
energy solutions of Euler equations has been proved before 
(\cite{s}, \cite{csy}). The particular type of solution we will describe
was proposed in (\cite{gfd}). The Eulerian-Lagrangian approach we take (\cite{el}) is not restricted to this particular case, but exact integration of the 
equations is.
We will consider a two dimensional basic square $Q$ of side $L$. The 
particular form of the solutions (\cite{gfd}) is 
\begin{equation}
{\mathbf u}(x,y,z,t) =\left (u(x,y,t), z\gamma(x,y,t)\right)
\label{boldu}
\end{equation}
where the scalar valued function $\gamma $ is periodic in both
spatial variables with period $L$ and
the two dimensional vector $u(x,y,t) = (u_1(x,y,t), u_2(x,y,t))$ is
also periodic with the same period. The associated two dimensional
curl is
\begin{equation}
\omega(x,y,t) = \frac{\partial u_2(x,y,t)}{\partial x} - \frac{\partial u_1(x,y,t)}{\partial y}.\label{omega}
\end{equation}
This represents the vertical (third) component of the vorticity 
$\nabla\times{\mathbf u}$ of the Euler
system and, using the ansatz (\ref{boldu}) it follows from the familiar
three dimensional vorticity equation that the equation
\begin{equation}
\partial_t\omega + u\cdot\nabla \omega = \gamma \omega
\label{omegaeq}
\end{equation}
should be satisfied for the recipe to succeed.
On the other hand, one can easily check that 
the vertical component of the velocity $z\gamma (x,y,t)$
solves the vertical component of the
Euler equations if $\gamma$ solves the non-local Riccati equation
\begin{equation}
\frac{\partial \gamma}{\partial t} + u\cdot\nabla \gamma
= -\gamma^2 + I(t)
\label{eq}
\end{equation}
where $I(t)$ is a function that depends on time only. 
The divergence-free condition for $\mathbf{u}$ becomes
\begin{equation}
\nabla\cdot u = - \gamma.\label{div}
\end{equation}
Because of the spatial periodicity of $u$ one must make sure that
\begin{equation}
\int_Q\gamma (x,t)dx = 0\label{cons}
\end{equation}
holds throughout the evolution. This can be done provided  
the function $I(t)$ is given by
\begin{equation}
I(t)  = \frac{2}{|Q|}\int_Q\gamma^2(x,t)dx\label{i}
\end{equation}
where
$$
|Q| = \int_Qdx = L^2.
$$ 
The velocity is determined from $\omega$ and 
$\gamma$ using a stream function $\psi (x,y,t)$
and a potential $h(x,y,t)$ by
\begin{equation}
u = \nabla^{\perp}\psi + \nabla h
\label{u}
\end{equation}
\begin{equation}
-\Delta h = \gamma\label{h},
\end{equation}
\begin{equation}
-\Delta \psi = \omega 
\label{curl}
\end{equation}
with periodic boundary conditions.

The ansatz ${\mathbf{u}}(x,y,z,t) =
(u(x,y,t), z\gamma(x,y,t))$ associates to
solutions of the system (\ref{omegaeq}, \ref{eq}, \ref{i}, \ref{u},
\ref{h}, \ref{curl}) in $d=2$ velocities
${\mathbf{u}}$ that obey the incompressible three dimensional
Euler equations (\cite {gfd}, \cite{og}).  The divergence
condition (\ref{div}) follows from (\ref{u}, \ref{h}). The compatibility
condition $\int_Q\omega dx =0$ is maintained
throughout the evolution because of $(\ref{omegaeq}, \ref{div})$.
We consider initial  data
\begin{equation}
\gamma (x,y, 0) = \gamma_0(x,y),\quad \omega(x,y, 0) = \omega_0(x,y)
\label{id}
\end{equation}
that are smooth and have mean zero
$\int_Q\gamma_0dx =\int_Q\omega_0dx = 0$. The solutions of the 
system above have local existence 
and the velocity is as smooth as the initial data are, as long as
$\int_0^T\sup_x|\gamma (x,t)|dt$ is finite (the result 
can be proved (\cite{gs}) following the idea of the proof of the 
well-known result (\cite{bkm})).
We will consider the characteristics
\begin{equation}
\frac{dX}{dt} = u(X,t)
\label{ch}
\end{equation}
and denoting $X(a,t)$ the characteristic that starts at $t=0$ from
$a$, $X(a,0) =a$, we note that, prior to blow up the map $a\mapsto X(a,t)$ is one-to-one
and onto as a map from ${\mathbf{T}}^2 = {\mathbf R}^2/L{\mathbf{Z}}^2$ to 
itself. The injectivity follows from the uniqueness of solutions of
ordinary differential equations. The surjectivity can be 
proved by reversing time on characteristics, which can be done
as long as the velocity is smooth. 
Our result is an explicit formula for $\gamma$ on characteristics
\begin{equation}
\gamma(x,t) = \alpha{(\tau(t))}\left\{\frac{\gamma_0(A(x,t))}{1 + \tau(t)\gamma_0(A(x,t))}
-\overline{\phi}(\tau(t))\right\}
\label{fora}
\end{equation}
where $A(x,t)$ is the inverse of $X(a,t)$ (the ``back-to-labels'' map)
and the functions $\tau(t)$, $\alpha(\tau)$ and $\overline{\phi}(\tau)$
are computed from the initial datum $\gamma_0$. More precisely
we show
\begin{thm}
Consider the nonlocal conservative Riccati system 
(\ref{omegaeq}, \ref{eq}, \ref{i}, \ref{u},
\ref{h}, \ref{curl}). There exist
smooth, mean zero initial data for which the solution becomes
infinite in finite time. Both the maximum and the minimum values of
the solution $\gamma$ diverge, to plus infinity
and respectively to negative infinity at the blow up time. 
There is no initial datum for which only the minimum diverges.
The general solution is given on characteristics in terms of 
the initial data $\gamma(x,0) = \gamma_0(x)$ by (see (\ref{forone}))
$$
\gamma(X(a,t), t) = \alpha(\tau(t))\left (\frac{\gamma_0(a)}{1+\tau(t)\gamma_0(a)}-\overline{\phi}(\tau(t))\right  )
$$
where
$$
\overline{\phi}(\tau) = \left\{\int_Q\frac{\gamma_0(a)}{(1 + \tau\gamma_{0}(a))^2}da\right\}
\left\{\int_Q\frac{1}{1 + \tau\gamma_{0}(a)}da\right\}^{-1},
$$
$$
\alpha(\tau) = \left\{\frac{1}{|Q|} \int_Q\frac{1}{1 + \tau\gamma_{0}(a)}da\right\}^{-2} 
$$
and
$$
\frac{d\tau}{dt} = \alpha(\tau),\quad \tau(0) = 0.
$$
The function $\tau(t)$ can be obtained from
$$
t = \left (\frac{1}{|Q|}\right )^2\int_Q\int_Q\frac{1}{\gamma_0(a)-\gamma_0(b)}
\log{\left(\frac{1+ \tau \gamma_0(a)}{1+\tau \gamma_0(b)}\right )}dadb.
$$
The Jacobian $J(a,t) = {\rm Det}\left\{\frac{\partial X(a,t)}{\partial a}\right\}$
is given by
$$
J(a,t) = \frac{1}{1 + \tau(t)\gamma_0(a)}\left\{\frac{1}{|Q|}\int_Q\frac{da}{1 + \tau(t)\gamma_0(a)}\right\}^{-1}
$$
The moments of $\gamma$ are given by
$$
\int_Q(\gamma(x,t))^pdx = (\alpha (\tau))^p\int_Q\left \{ \frac{\gamma_0(a)}{1+\tau(t)\gamma_0(a)}-\overline{\phi}(\tau(t))\right\}^{p}J(a,t)da.
$$
The blow up time $t =T_*$ is given by
$$
T_* =\frac{1}{|Q|^2}\int\int\frac{1}{\gamma_0(a)-\gamma_0(b)}
\log{\left (\frac{\gamma_0(a)-m_0}{\gamma_0(b)- m_0}\right)}dadb
$$
where
$$
m_0 = \min_Q{\gamma_0(a)}<0.
$$
\end{thm}

We note that the curl $\omega$ 
plays a secondary role in this calculation and in the blow up. Indeed, 
the same formula and blow up occurs if $\omega_0 =0$, or if the curl
$\omega$ was smooth and computed in a different fashion 
than via (\ref{omegaeq}).

\section{Solving on characteristics}
We will solve now the nonlocal Riccati equation on characteristics.
We start with an auxiliary problem. Let
$\phi$ solve
$$
\partial_\tau\phi  + v\cdot\nabla \phi = -\phi^2
$$
together with
$$
\nabla\cdot v(x,\tau ) = -\phi(x,\tau ) + \frac{1}{|Q|}\int_Q\phi(x,\tau)dx.
$$
We will consider initial data that are smooth, periodic and
have zero mean,
$$
\int_Q\phi_0(x)dx = 0.
$$
We will also assume that the curl $\zeta = \frac{\partial v_2}{\partial x} -
\frac{\partial v_1}{\partial y}$ obeys
$$
\partial_\tau \zeta + v\cdot\nabla \zeta = 
\left (\phi -\frac{3}{|Q|}\int_Q\phi(x,\tau)dx \right )\zeta . 
$$ 
Passing to characteristics
\begin{equation}
\frac{dY}{d\tau} = v(Y,\tau)
\label{charphi}
\end{equation}
we integrate and obtain
$$
\phi(Y(a,\tau),\tau ) = \frac{\phi_0(a)}{1+ \tau\phi_0(a)}
$$
valid as long 
$$
\inf_{a\in Q}(1+\tau\phi_0(a)) > 0.
$$
We need to compute 
$$
\overline{\phi}(\tau ) = \frac{1}{|Q|}\int_Q\phi(x,\tau)dx.
$$
The Jacobian 
$$
J(a,\tau) = Det\left \{\frac{\partial Y}{\partial a}\right \}
$$
obeys
$$
\frac{dJ}{d\tau } = -h(a,\tau )J(a,\tau)
$$
where
$$
h(a,\tau ) = \phi(Y(a,\tau ),\tau) -\overline{\phi}(\tau).
$$ 
Initially the Jacobian equals to one, so
$$
J(a, \tau) = e^{-\int_0^{\tau} h(a,s)ds}.
$$
So
$$
J(a,\tau) = e^{\int_0^{\tau}\overline{\phi}(s)ds}\exp{(-\int_0^{\tau}\frac{d}{ds}\log(1+s\phi_0(a))ds)}
$$
and thus
$$
J(a,\tau) = e^{\int_0^{\tau}\overline{\phi}(s)ds}\frac{1}{1+\tau\phi_0(a)}.
$$
The map $a\mapsto Y(a,\tau)$ is one and onto. The change of variables 
formula gives
$$
\int_Q\phi(x,\tau)dx = \int_Q \phi(Y(a,\tau),t)J(a, \tau)da
$$
and therefore
\begin{equation}
\overline{\phi}(\tau) = e^{\int_0^{\tau}\overline{\phi}(s)ds}\frac{1}{|Q|}
\int_Q\frac{\phi_0(a)}{(1 + \tau\phi_{0}(a))^2}da.
\label{ex}
\end{equation}
Consequently 
$$
\frac{d}{d\tau}e^{-\int_0^{\tau}\overline{\phi}(s)ds} = \frac{d}{d\tau}\frac{1}{|Q|}
\int_Q\frac{1}{1 + \tau\phi_{0}(a)}da.
$$
Because both sides at $\tau = 0$ equal one, we have
\begin{equation}
e^{-\int_0^{\tau}\overline{\phi}(s)ds} = 
\frac{1}{|Q|}\int_Q\frac{1}{1 + \tau\phi_{0}(a)}da\label{ephibar}
\end{equation} 
and, using (\ref{ex}),
\begin{equation}
\overline{\phi}(\tau) = 
\left\{\int_Q\frac{\phi_0(a)}{(1 + \tau\phi_{0}(a))^2}da\right\}
\left\{\int_Q\frac{1}{1 + \tau\phi_{0}(a)}da\right\}^{-1}
\label{phibar}
\end{equation}

Note that the function $\delta(x,\tau) = \phi(x,\tau ) -\overline{\phi}(\tau )$ 
obeys
$$
\frac{\partial \delta}{\partial \tau } + v\cdot\nabla \delta =
-\delta^2 + 2\frac{1}{|Q|}\int_Q\delta ^2dx -2\overline{\phi}\delta.
$$
We consider now the function
$$
\sigma(x,\tau ) = e^{2\int_0^{\tau}\overline{\phi}(s)ds}\delta(x,\tau)
$$
and the velocity
$$
U(x,\tau ) =  e^{2\int_0^{\tau}\overline{\phi}(s)ds}v(x,\tau).
$$
Multiplying the equation of $\delta$ by $ e^{4\int_0^{\tau}\overline{\phi}(s)ds}$
we obtain
$$
e^{2\int_0^{\tau}\overline{\phi}(s)ds}\frac{\partial \sigma }{\partial {\tau}}
+ U\cdot\nabla\sigma = -\sigma^2 + \frac{2}{|Q|}\int{\sigma^2dx}.
$$
Note that
$$
\nabla\cdot U = - \sigma.
$$
Now we change the time scale.
We define a new time $t$ by the equation
\begin{equation}
\frac{dt}{d\tau } = e^{-2\int_0^{\tau}\overline{\phi}(s)ds},
\label{timeq}
\end{equation}
$t(0) = 0$, and new variables
$$
\gamma(x,t) = \sigma(x,\tau)
$$
and
$$
u(x,t) = U(x,\tau).
$$
Now $\gamma$ solves the nonlocal conservative Riccati equation
\begin{equation}
\frac{\partial \gamma}{\partial t} + u \cdot\nabla \gamma =
-\gamma^2 + \frac{2}{|Q|}\int{\gamma^2dx}\label{ric}
\end{equation}
with periodic boundary conditions, 
\begin{equation}
u = (-\Delta)^{-1}\left [\nabla^{\perp}\omega + \nabla\gamma \right ]
\label{eqns}
\end{equation}
and 
\begin{equation}
\frac{\partial \omega }{\partial t} + u\cdot\nabla\omega = \gamma \omega
\label{om}
\end{equation}
The initial data are given by
$$
\gamma_0(x) = \delta_0(x) = \phi_0(x).
$$
Using (\ref{ephibar}) and integrating
the equation (\ref{timeq})
we see that the time change is given by the formula
\begin{equation}
t = \left (\frac{1}{|Q|}\right )^2\int_Q\int_Q\frac{1}{\phi_0(a)-\phi_0(b)}\log{\frac{1+\tau \phi_0(a)}{1+\tau \phi_0(b)}}dadb
\label{time}
\end{equation}
Note that the characteristic system
$$
\frac{dX}{dt} = u(X,t)
$$
is solved by
$$
X(a,t) = Y(a,\tau)
$$
where $Y$ solves the system (\ref{charphi}).
This implies the formula
\begin{equation}
\gamma(X(a,t),t) = \alpha (\tau)\left (
\frac{\phi_0(a)}{1+\tau \phi_0(a)}-\overline{\phi}(\tau)\right ) 
\label{forone}
\end{equation}
with
\begin{equation}
\alpha (\tau) =  e^{2\int_0^{\tau}\overline{\phi}(s)ds}.
\label{alpha}
\end{equation}
In view of (\ref{ephibar}), (\ref{phibar}), (\ref{time}),  we have obtained a 
complete description of the general solution in terms of the initial 
data. 
\section{Blow up}

Consider an initial
smooth function $\gamma_0(a) = \phi_0(a) $ and assume that it has mean zero and that its 
minimum is $m_0<0$. As it is
evident from the explicit formula the blow up time for $\phi(Y(a,\tau),\tau)$ is
$$
\tau_* = -\frac{1}{m_0}
$$ 
and $\phi(Y(a,\tau),\tau)$ diverges to negative infinity for some $a$, and
not at all for others. This of course does not necessarily mean that $\gamma$ blows up in the same 
fashion. Let us discuss the simplest case, in 
which the minimum is attained at a finite number of locations $a_0$, and
near these locations, the function $\phi_0$ has non-vanishing second 
derivatives, so that locally 
$$
\phi_0(a) \ge m_0 + C|a-a_0|^2
$$
for $0\le |a-a_0|\le r$. 
Then it follows that the integral
$$
\frac{1}{|Q|}\int\frac{da}{\epsilon^2 + \phi_0(a)-m_0}
$$
behaves like
$$
\frac{1}{|Q|}\int\frac{da}{\epsilon^2 + \phi_0(a)-m_0}\sim \log\left \{\sqrt{1+\left (\frac{Cr}{\epsilon}\right )^2}\right \}
$$
for small $\epsilon$. Taking
$$
\epsilon^2 = \frac{1}{\tau } - \frac{1}{\tau_*}  
$$
we deduce that, for these kinds of initial data
$$
e^{-\int_0^{\tau}\overline{\phi}(s)ds} \sim 
\log \left \{\sqrt{1+\frac{C}{\tau_*-\tau}}\right \}.
$$
For the same kind of functions and small $(\tau_*-\tau)$, the integral
$$
\frac{1}{|Q|}\int_Q\frac{\phi_0(a)}{(1+\tau\phi_0(a))^2}da
\sim -\frac{C}{\tau_*-\tau}
$$
and $t(\tau)$ has a finite limit $t\to T_*$
as $\tau \to \tau_*$. The average $\overline{\phi}(\tau )$ diverges to negative
infinity,
$$
\overline{\phi}(\tau) \sim -\frac{C}{\tau_*-\tau}\left [\log \left \{\sqrt{1+\frac{C}{\tau_*-\tau}}\right \}\right ]^{-1}.
$$
The prefactor $\alpha$ becomes
vanishingly small
$$
\alpha (\tau) \sim (\log (\tau_*-\tau))^{-2}
$$
and  (\ref{forone}) becomes 
$$
\gamma (X(a,t),t) \sim (\log (\tau_*-\tau))^{-2}\left (\frac{\phi_0(a)}{1+\tau\phi_0(a)}-\overline{\phi}(\tau )\right).
$$
If the label is chosen so that
$\phi_0(a)>0$ then the first term in the brackets does not blow up
and $\gamma$ diverges to plus infinity. If the label is chosen at
the minimum, or nearby, then the first term in the brackets
dominates and the blow up is to negative infinity, as expected from the
ODE. From the equation (\ref{timeq})
$$
(\alpha (\tau))^{-1}d\tau = dt
$$
it follows that
$$
T_*-t \sim (\tau_*-\tau)\left (1 + \log\left (\frac{1}{\tau_*-\tau}\right )\right)^2
$$
and the asymptotic behavior of the blow up in $t$ follows from the one in $\tau$. We end by addressing a question that was at some point raised by
numerical simulations: can there be a one-sided blow up? From the representation (\ref{forone}) 
of the solution it follows that
$$
M(t) \ge -\overline{\phi}(\tau)e^{2\int_0^{\tau}\overline{\phi}(s)ds}
$$
holds for $M(t)= \sup_x\gamma(x,t)$. If one would assume that, up to the
putative blow up
$$
M(t)\le C
$$
with some fixed constant $C$, then it would follow that
$$
-\frac{d}{d\tau }e^{2\int_0^{\tau}\overline{\phi}(s)ds} \le 2C
$$
and integrating between $\tau $ and $\tau_*$ that
$$
e^{2\int_0^{\tau}\overline{\phi}(s)ds} \le 2C(\tau_*-\tau).
$$
This in turn would imply that $T_* = \infty$ and therefore no blow up
for $\gamma$ can occur in finite $t$. So the answer is that for
no initial datum can there exist a one sided blow up for $\gamma$.

\noindent {\bf Acknowledgments} I thank J. Gibbon and K. Ohkitani
for showing me their numerical work prior to its publication. 
This work was partially supported by NSF-DMS9802611 and by
the ASCI Flash Center at the University of Chicago under DOE contract 
B341495.

\end{document}